\documentclass[letterpaper,twocolumn,10pt]{article}
\usepackage{zhanggroup}

\usepackage{algorithm}
\usepackage{algorithmic}
\usepackage{graphicx}
\usepackage{amsmath}
\usepackage{amssymb}
\usepackage{booktabs}
\usepackage{multirow}
\usepackage{times}
\usepackage{epsfig}
\usepackage{subcaption}
\usepackage{newfloat}
\usepackage{listings}

\newcommand{\mypara}[1]{\smallskip\noindent\textbf{#1.}}

\newcommand{\TargetEncoder}{{E}_\texttt{t}}
\newcommand{\TargetDecoder}{{D}_\texttt{t}}

\newcommand{\ShadowDecoder}{{D}_\texttt{s}}

\newcommand{\ShadowDataset}{\mathcal{D}_{\textit{shadow}}}

\newcommand{\TargetTrain}{\mathcal{D}_{\textit{target}}^{\textit{train}}}
\newcommand{\TargetTest}{\mathcal{D}_{\textit{target}}^{\textit{test}}}
\newcommand{\ShadowTrain}{\mathcal{D}_{\textit{shadow}}^{\textit{train}}}
\newcommand{\ShadowTest}{\mathcal{D}_{\textit{shadow}}^{\textit{test}}}
\usepackage[capitalize]{cleveref}
\crefname{section}{Sec.}{Secs.}
\Crefname{section}{Section}{Sections}
\Crefname{table}{Table}{Tables}
\crefname{table}{Tab.}{Tabs.}

\title{\bf Membership Inference Attack Against Masked Image Modeling}

\author{
Zheng Li\textsuperscript{1}\ \ \
Xinlei He\textsuperscript{1}\ \ \
Ning Yu\textsuperscript{2}\ \ \
Yang Zhang\textsuperscript{1}\ \ \
\\
\\
\textsuperscript{1}\textit{CISPA Helmholtz Center for Information Security} \ \ \ 
\textsuperscript{2}\textit{Netflix Eyeline Studios}
}

\begin{document}
\date{}
\maketitle

\begin{abstract}
Masked Image Modeling (MIM) has achieved significant success in the realm of self-supervised learning (SSL) for visual recognition. 
The image encoder pre-trained through MIM, involving the masking and subsequent reconstruction of input images, attains state-of-the-art performance in various downstream vision tasks. 
However, most existing works focus on improving the performance of MIM. 
In this work, we take a different angle by studying the pre-training data privacy of MIM. 
Specifically, we propose the first membership inference attack against image encoders pre-trained by MIM, which aims to determine whether an image is part of the MIM pre-training dataset.
The key design is to simulate the pre-training paradigm of MIM, i.e., image masking and subsequent reconstruction, and then obtain reconstruction errors.
These reconstruction errors can serve as membership signals for achieving attack goals, as the encoder is more capable of reconstructing the input image in its training set with lower errors.
Extensive evaluations are conducted on three model architectures and three benchmark datasets. 
Empirical results show that our attack outperforms baseline methods. 
Additionally, we undertake intricate ablation studies to analyze multiple factors that could influence the performance of the attack.
\end{abstract}

\section{Introduction}\label{sec:intro}
Machine Learning (ML) has demonstrated remarkable advancements across critical domains, including face recognition and natural language processing. 
Contemporary ML models demand vast amounts of labeled data, often inaccessible to the public, and exhibit susceptibility to data overfitting. 
This data requirement has recently found resolution through a prominent technique known as contrastive self-supervised learning (SSL)~\cite{ZZSYZK22,CKNH20,OLV18,RKHRGASAMCKS21}. 
Further, with the emergence of Transformer architecture, Masked Image Modeling (MIM) has attracted unprecedented attention for achieving state-of-the-art performance in vision tasks~\cite{ABCGL22, BMAZ22,HCXLDG21,BDW21}, and has been applied in various application, such as video~\cite{WFXWYF21, TSWW22}, graph~\cite{TLHCCH22} and audio~\cite{BPH22}.

\begin{figure}[t]
    \centering
    \includegraphics[width=0.9\linewidth]{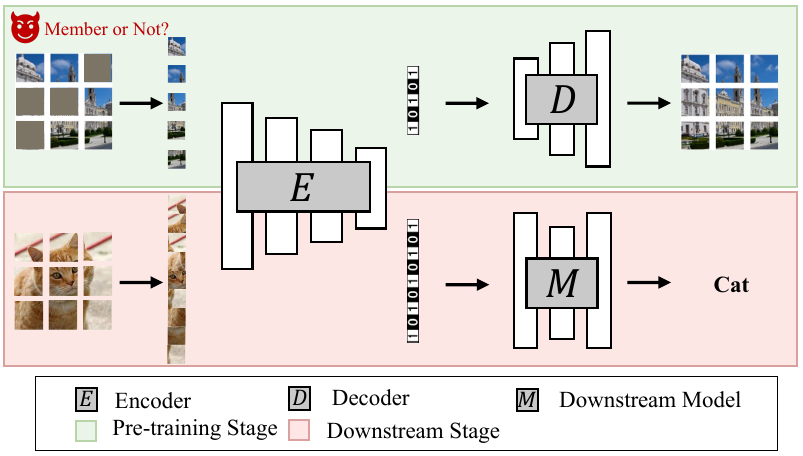}
    \caption{An illustration of the pre-training stage and downstream stage of Masked Image Modeling (MIM).}
    \label{intro}
\end{figure}

Unlike SSL with an encoder aligning different augmented views of the same image, MIM has an asymmetric encoder decoder design, where the encoder randomly masks patches of the input image, and the decoder reconstructs the missing patches based on the previous-stage unmasked patches.
Given an unlabeled image dataset (called pre-training dataset), MIM pre-trains an image encoder by the masking and reconstructing workflow, and then the well-pre-trained image encoder can be used as a feature extractor for various downstream tasks.
See Figure \ref{intro} for an illustration of the pre-training stage and downstream stage.
Typically, the pre-training of encoders demands huge computational resources and millions of unlabeled data.
Thus, a powerful encoder provider, e.g., Meta, Google, and OpenAI, pre-trains encoders by MIM technique and provides these encoders to the public or specific downstream customers who lack computational resources or data.

\subsection{Motivation}
Despite its remarkable performance, MIM also relies on extensive (albeit unlabeled) large-scale data for its construction. This large-scale dataset typically encompasses sensitive and private individual information, including health conditions and social relationships. 
Recent studies~\cite{SSSS17, SZHBFB19, HYYBGC21,ONK21, HWWBSZ21, WYPY21} have highlighted the susceptibility of supervised learning and contrastive SSL to privacy attacks, specifically through membership inference attacks~\cite{SSSS17}: An attacker seeks to determine whether a given sample belongs to the training set of the target model. 
Nevertheless, the current development of MIM only focuses on enhancing its performance and expanding its application in more tasks. 
The privacy risks stemming from them are largely unexplored.

\subsection{Contributions}
In this work, we take the first step towards studying the pre-training data privacy of MIM through the lens of membership leakage.
Specifically, we aim to infer whether or not a given image is part of the pre-training dataset of MIM (see Figure \ref{intro}).
If the image is in the pre-training dataset, we call it a member, otherwise, a non-member.
To this end, we introduce a new attack methodology tailored for MIM. 
Given a target image and a pre-trained image encoder, our attack initiates by emulating the masking and reconstruction paradigm of MIM. 
Then, we quantify the distance between the reconstructed and target images, categorizing it as a member if the distance falls below a threshold, and as a non-member otherwise. 
We conduct extensive experiments on multiple model architectures and benchmark datasets. 
Empirical results demonstrate that our attack achieves notably superior performance, and existing attacks cannot be extended to MIM.
This underscores the necessity of our attack specifically designed for MIM. 
Furthermore, we undertake a comprehensive ablation study, exploring various aspects of MIM that could potentially influence the attack performance. 
Lastly, we delve into a further investigation of our attack by relaxing several adversary's assumptions.
Our main contributions are as follows:
\begin{itemize}
    \item We take the first setup to study the pre-training data privacy of MIM through the lens of membership leakage.
    \item We propose the first uniquely designed membership inference attack against MIM, which is initiated by emulating the masking and reconstruction paradigm of MIM. 
    \item We conduct extensive evaluation and demonstrate that our attack can achieve superior performance than existing works.
    \item We perform sophisticated ablation studies by analyzing different aspects and relaxing some assumptions of the adversary to study their impact on the attack performance.
\end{itemize}

\section{Background}

\subsection{Membership Inference Attacks}
Membership inference attacks~\cite{SSSS17, SZHBFB19, HYYBGC21,ONK21, HWWBSZ21, WYPY21,LLHYBZ22,HLXCZ22,WYLBZ22, LLWHYZFZ24, CHLZL23, XLYCFBZ24, HSLGBZ24} aim to identify whether a target data sample exists within the training set of the given model. 
Successful membership inference attacks can have several consequences: (1) raising privacy concerns due to their potential to expose sensitive information, such as health conditions; (2) enabling adversaries to access additional information about the model's dataset, thereby infringing upon the model's intellectual property; (3) auditing whether their data is being utilized by the model.

Within the domain's literature, Shokri et al.~\cite{SSSS17} introduce the pioneering score-based membership inference attack against supervised learning, achieved through the construction of multiple shadow and attack models. 
In addition to diverse score-based attacks~\cite{SZHBFB19, HYYBGC21,ONK21, HWWBSZ21, WYPY21}, Li and Zhang~\cite{LZ21} and Choo et al.~\cite{CTCP21} explore a more strict scenario in which the adversary solely possesses access to the predicted labels generated by the given model. 
In the realm of generative modeling, Chen et al.~\cite{CYZF20}, and Hilprecht et al.~\cite{HHB19} assess the probability that a target sample can be generated by a generator, based on latent codes that lead to image outputs.
This estimation prioritizes members' probability over that of non-members.

However, prevailing and extensively studied attack methodologies cannot be readily applied to MIM, as existing works mainly focus on supervised learning and generative modeling, while MIM enjoys a new and unique training paradigm, i.e., masking and reconstructing random patches of the input image.

\subsection{Masked Image Modeling (MIM)}
Inspired by the success of Masked Language Modeling (MLM)~\cite{DCLT19,XTLYL20,ZXGMXW20,HHSJCL20} in the pre-training NLP domain, MIM~\cite{HCXLDG21,XZCLBYDH21,CDWXMWHLZW22,WFXWYF21,ZZSYZK22} has been recently proposed and demonstrated the great potential in pre-training vision domain.
MIM follows an asymmetric encoder-decoder design with transformer architecture, and its key idea is to mask random patches of the input image and reconstruct the lost patches.
Concretely, the input image is first divided into regular non-overlapping patches, and the encoder randomly masks some patches. 
Then the encoder maps the remaining unmasked patches to a latent representation, and a lightweight decoder reconstructs the original image from latent representations.
The encoder and decoder are optimized by minimizing the distance between predicted patches and masked patches. 

To adapt the pre-trained encoder to various downstream tasks, the downstream customers first build the downstream model by adding only a few task-specific layers on top of the pre-trained encoder.
Then the downstream customers can use two adaptation approaches, namely \textit{end-to-end fine-tuning} and \textit{linear probing}.
The former means updating the entire model, while linear probing means updating only the newly added additional layers.
Almost all the existing works~\cite{HCXLDG21,XZCLBYDH21,CDWXMWHLZW22,WFXWYF21} show that the end-to-end fine-tuning of the downstream model is significantly superior to the linear probing, and even exceeding 10\%~\cite{HCXLDG21,BDW21}, which is a performance gain that cannot be ignored.
Therefore, we argue that the most realistic and widespread scenario is that \textit{the encoder provider releases the pre-trained encoder to the public or specific downstream customers for full access.}
This realistic scenario further forms the basis of our threat model, in which the adversary can fully access the pre-trained encoder.

\section{Membership Inference Attack Against MIM}
We start by introducing the threat model. 
Then we introduce our key intuition.
Finally, we describe the attack methodology.

\subsection{Threat Model}\label{sec:threat_model}
\mypara{Adversary's Goal}
Given an input image $x$ (called target image) and an image encoder (called target encoder $\TargetEncoder$) pre-trained by MIM, an inferrer aims to infer whether or not the target image $x$ is in the pre-training set.
We call the target image a \textit{member} if it is in the pre-training set of the target encoder. Otherwise, we call it a \textit{non-member}.

\mypara{Adversary's Knowledge}
We consider the adversary to have full access to the target encoder $\TargetEncoder$.
We emphasize that this is the most realistic and widespread scenario due to the superiority of fine-tuning entire downstream models over linear probing.
Thus, pre-trained encoders should be made accessible to the public or downstream customers.
A typical scenario is that the encoder provider pre-trains an asymmetric encoder-decoder structure through MIM, and only provides the pre-trained encoder to downstream customers for fine-tuning.
Then the downstream customer, actually disguised by an adversary, aims to infer sensitive or private information or detect whether his/her public data is abused without his/her authorization.
Note that here, we assume that the adversary can only access the target encoder and not its associated decoder $\TargetDecoder$.
In addition, we assume that the adversary knows the exact mask ratio adopted by the target encoder.

Further, we assume the adversary has access to a local dataset (called \textit{shadow dataset}) that has the same distribution as the pre-training dataset.
Here we emphasize that almost all existing work in the area of member inference attacks makes the same assumptions and we follow them~\cite{SZHBFB19, HYYBGC21,ONK21, HWWBSZ21, WYPY21, CTCP21, LZ21}.
Note that we further relax this assumption in the experiments.

\subsection{Key Intuition}
The key intuition behind our attack is the training paradigm of MIM.
Specifically, the encoder and decoder are optimized to fit their pre-training dataset, so that the reconstructed image should be closer (or less) to the image in (or not) the pre-training dataset.
That is, members and non-members actually behave differently regarding the distance between the reconstructed image and the input image.
Thus, the key design of our attack is to simulate the training paradigm of MIM, i.e., masking random patches of the input image and reconstructing missing patches.
We then measure the distance between the reconstructed image and the input image and treat it as a member if the distance is less than a threshold.

\mypara{Changes} In summary, the main challenges are as follows:
\begin{itemize}
    \item How to establish an optimal threshold that effectively discriminates between members and non-members.
    \item How to simulate the masking and reconstruction paradigm when only the target encoder is available, without its corresponding decoder?
\end{itemize}

\begin{figure}[t]
    \centering
    \includegraphics[width=0.9\linewidth]{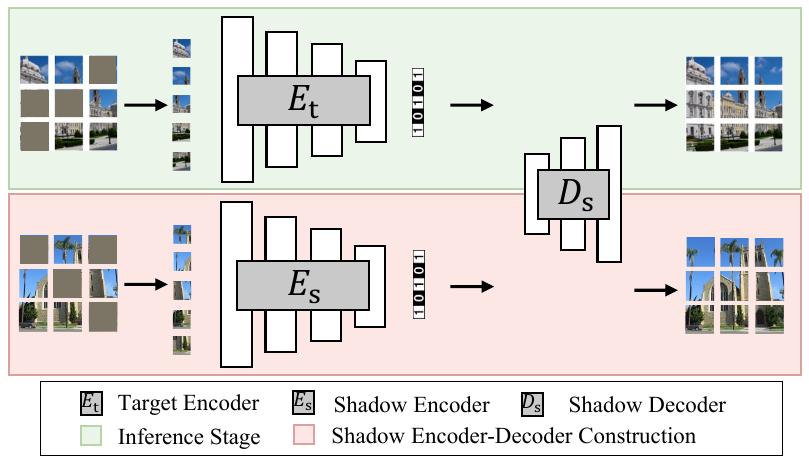}
    \caption{Overview of our attack mechanism against Masked Image Modeling (MIM).}
    \label{method}
\end{figure}

\subsection{Attack Methodology}
To tackle these challenges, we present a new membership inference attack designed specifically for MIM.
The attack methodology can be divided into three main stages, namely Threshold Search, Target Encoder-Decoder Simulation, and Membership Inference.
Figure \ref{method} shows the overview of our attack.

\mypara{Threshold Search}
To find the threshold, the adversary first splits his/her local dataset (i.e., $\ShadowDataset$) into two disjoint sets, $\ShadowTrain$ (member) and $\ShadowTest$ (non-member).
The adversary can then use $\ShadowTrain$ to train the local encoder and decoder via the MIM training paradigm.
The adversary feeds members and non-members into the trained shadow encoder-decoder to produce their respective reconstructed images.
Then, the adversary measures the distance between the reconstructed images and the input image for members and non-members, respectively.
Finally, the adversary can plot ROC curves based on various thresholds, and finally obtain the optimal threshold $t$ at which the shadow encoder can perfectly distinguish members from non-members.

Regarding the shadow encoder architecture, the adversary can construct it by directly utilizing the target encoder, since the adversary has full access to the target encoder.
For the shadow decoder architecture, we use the same architecture as the target decoder, which is also adopted by existing works~\cite{SSSS17, SZHBFB19, WYPY21, CTCP21, LZ21}.
We here emphasize that the adversary does not have the knowledge of the target decoder architecture, and in the following experiments, we show that a wide range of shadow decoder architecture choices yield similar attack performance.

\mypara{Target Encoder-Decoder Simulation}
Once the threshold is obtained, the adversary uses the trained shadow decoder $\ShadowDecoder$ as the target encoder by sequentially connecting it to the target encoder $\TargetEncoder$.
The adversary then only fine-tunes the shadow decoder $\ShadowDecoder$ by feeding $\ShadowTrain$ to the simulated target encoder-decoder, guiding the shadow decoder closer to the target decoder.
In this way, the adversary can successfully simulate the target encoder-decoder.

\mypara{Membership Inference}
Given a target image $x$ from $\TargetTrain$ or $\TargetTest$, the adversary queries it directly to the simulated target encoder-decoder to mimic the masking and reconstruction paradigms of MIM.
The adversary can then obtain the reconstructed image and measure the distance between the reconstructed and target images.
Finally, the adversary simply considers the target images whose reconstruction distances are less than the searched threshold $t$ as members and vice versa.

\section{Experimental Setup}
\subsection{Target Encoder-Decoder}
We evaluate our attack methodology using an open-source implementation of MIM~\cite{HCXLDG21}, which is the most representative and has attracted unprecedented attention.
Typically, the target encoder in MIM is based on a standard vision transformer (ViT) ~\cite{DBKWZUDMHGUH21} with only \textit{visible, unmasked patches} being fed to the ViT to learn the representation.
Thus, without losing representativeness, we consider target encoders with different complexities of ViT architectures, i.e., ViT-Base/Large/Huge, as shown in Table \ref{tab:model_arches}.
Besides, we adopt the optimal mark ratio for the best performance, i.e., 0.75, as well as 500 epochs to pre-train the encoder-decoder.

For the target decoder, we follow the original implementation using a series of Transformer blocks, i.e., a lightweight Transformer with 8 layers.
Note that the decoder is only used to perform the image reconstruction task during the pre-training period.
Therefore, the decoder architecture can be flexibly designed independently of the encoder design.

\subsection{Datasets}
We utilize three benchmark datasets in our experiments, i.e., CIFAR-10~\cite{krizhevsky2009learning}, CIFAR-100~\cite{krizhevsky2009learning}, and TinyImageNet~\cite{le2015tiny}.
All images are resized to $224 \times 224$ to fit the input requirement of the models, which is also a common practice in related works~\cite{JLG22, DBKWZUDMHGUH21}.
All three encoder-decoder pairs fit these three datasets well, and we show their fast-decreasing training losses in Appendix Figure \ref{fig:training_loss}, which verifies the correctness of our training setup.
The reason we do not adopt ImageNet is that the target encoder shown in Table \ref{tab:model_arches} contains about 86M/307M/632M parameters.
Training this encoder on ImageNet would demand an extensive time investment and substantial computing resources, particularly considering that ImageNet encompasses 1000 classes and over 1 million images.

\subsection{Metric}
We adopt the attack success rate (denoted as \textit{ASR}) to evaluate attack performance. 
Given an evaluation dataset that contains ground truth members and non-members of the target encoder, \textit{ASR} is the ratio of the ground truth members/non-members that are correctly differentiated.

\subsection{Baseline Attacks}
Existing membership inference attacks aim to infer members of a classifier or an image generator, and these existing attacks cannot be trivially extended to MIM that enjoys a new and unique training paradigm.
We generalize these existing attacks to MIM as baseline attacks. 
In particular, we compare our attack with the following four baseline attacks.
\begin{table}[!t]
    \centering
    \caption{Target encoder with different complexities of ViT architectures, i.e., ViT-Base/Large/Huge, as well as target decoder with a few Transformer blocks.}
    \scalebox{0.80}
    {
    \begin{tabular}{c|c|c|c}
    \toprule
        Target Encoder-Decoder & Base Architecture & Layers & Params\\
         \midrule
         \textsc{Encoder-I} & ViT-Base & 12 & 86M\\
         \midrule
          \textsc{Encoder-II} & ViT-Large & 24  & 307M\\
         \midrule
         \textsc{Encoder-III} & ViT-Huge & 32  & 632M\\
         \midrule
         \textsc{Decoder} & Transformer & 8  & 23M\\
         \bottomrule
    \end{tabular}
    }
    \label{tab:model_arches}
\end{table}

\mypara{Baseline-A}
In this baseline, the adversary first builds a downstream classifier (called \textit{target downstream classifier}) for a downstream task based on the target encoder and then leverages existing membership inference methods~\cite{SSSS17, SZHBFB19} to attack the new downstream classifier. 
Specifically, we consider CINIC-10~\cite{DCAS18} as a downstream task and randomly select 20,000 training samples as the downstream dataset.
Similarly, the adversary trains a \textit{shadow downstream classifier} using the downstream dataset.
Finally, the adversary can train a binary classifier based on the output posterior of the shadow downstream classifier by feeding the members and non-members of its shadow pre-training dataset, and utilize the well-trained binary classifier to infer the members of the target encoder based on output posterior of the target downstream classifier.

\mypara{Baseline-B} 
In this baseline, the adversary treats the target encoder directly as a classifier and launches the white-box membership inference attack that exploits both the output embedding and internal features of the target encoder.
The adversary first trains a shadow encoder, and then trains a binary classifier based on the output embedding and internal features of the shadow encoder by feeding the members and non-members of its shadow dataset.
Finally, the adversary utilizes the well-trained binary classifier to infer the members of the target encoder based on the output embedding and internal features of the target encoder.

\begin{figure*}[!t]
\centering
\begin{subfigure}{0.6\columnwidth}
\includegraphics[width=\columnwidth]{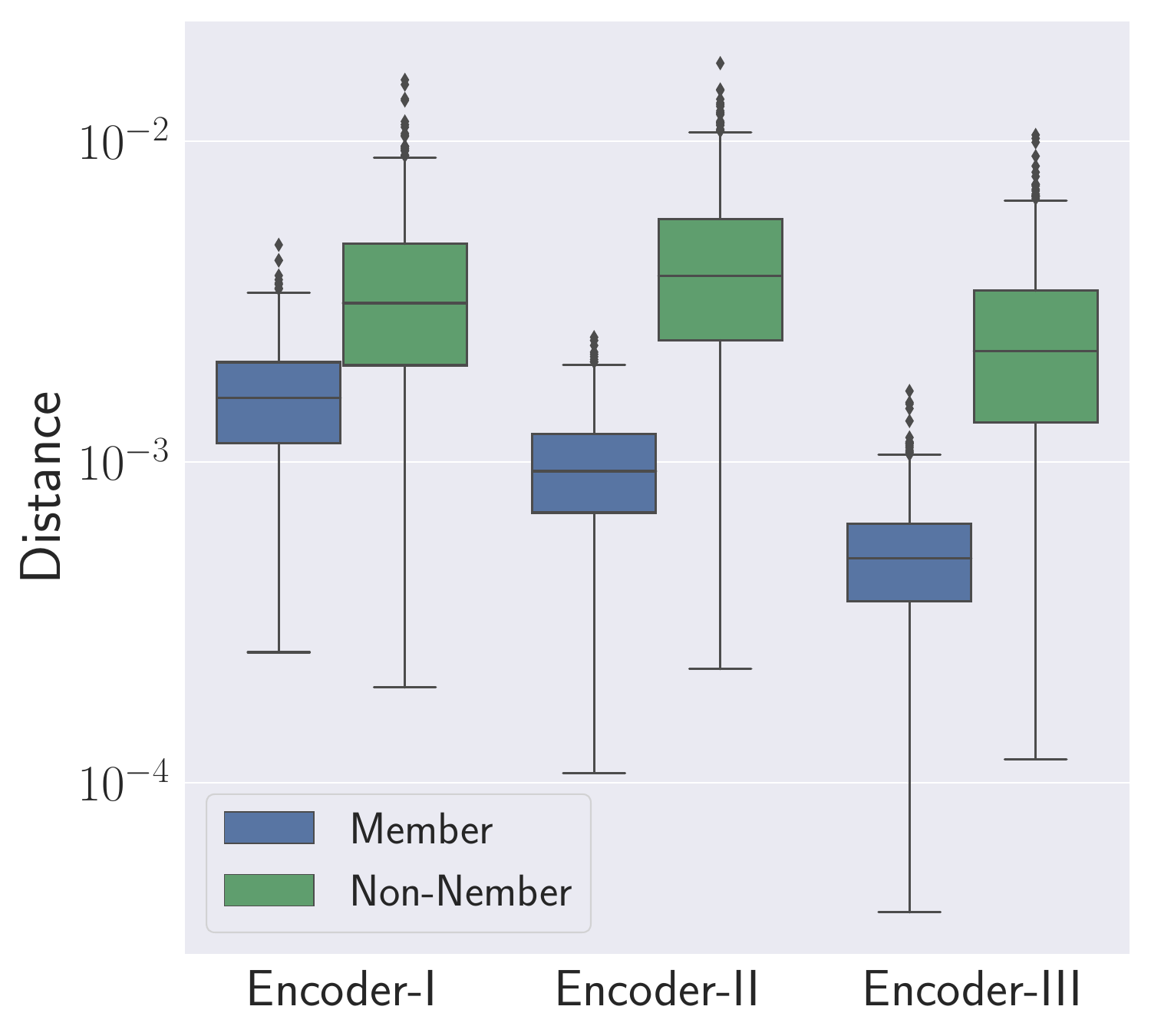}
\caption{CIFAR-10}
\end{subfigure}
\begin{subfigure}{0.6\columnwidth}
\includegraphics[width=\columnwidth]{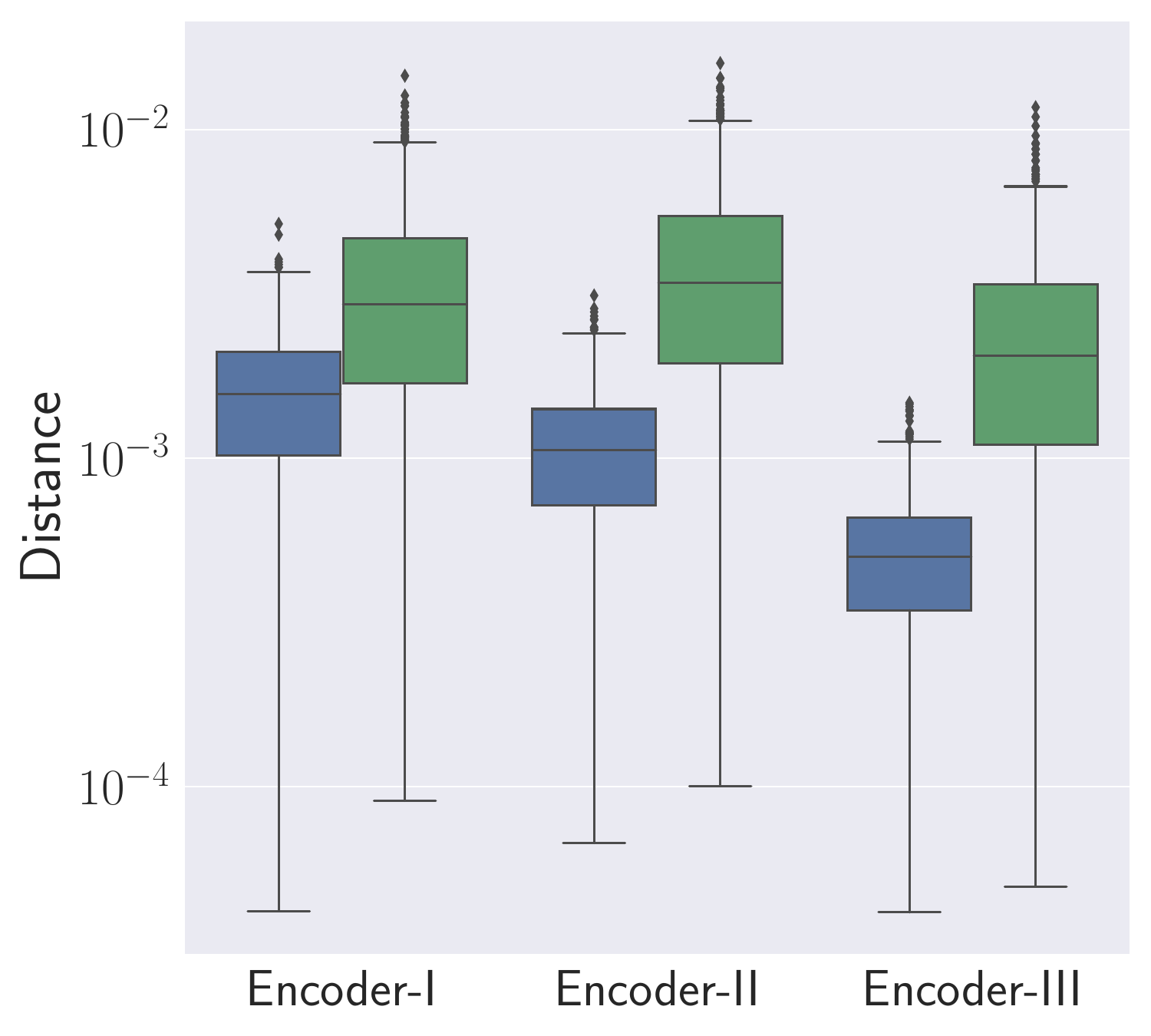}
\caption{CIFAR-100}
\end{subfigure} 
\begin{subfigure}{0.6\columnwidth}
\includegraphics[width=\columnwidth]{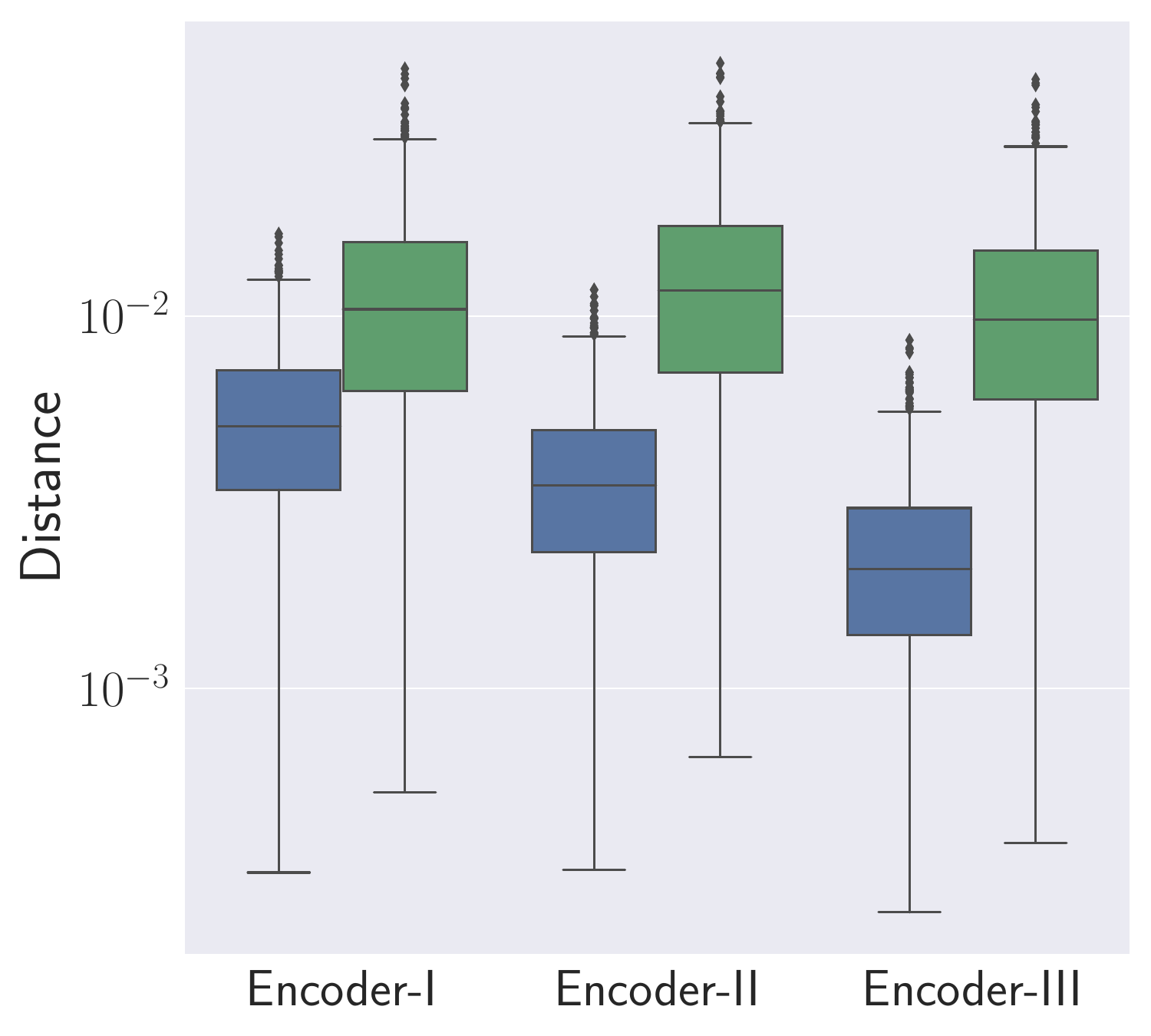}
\caption{TinyImageNet}
\end{subfigure}
\caption{The distribution of distance between the reconstructed images and original images for members and non-members.}
\label{fig:reconstructed_distance}
\end{figure*}

\begin{table*}[!t]
\centering
\caption{Attack performance of baseline attacks and our proposed attack.}
\scalebox{0.85}{
\begin{tabular}{l|l|c|c|c|c|c}
\toprule
\multirow{2}{*}{Dataset}      & \multirow{2}{*}{Encoder} & \multicolumn{5}{c}{Membership Inference Attack}          \\
                              &                          & Baseline-A & Baseline-B & Baseline-C & Baseline-D & Our Attack \\
\midrule
\multirow{3}{*}{CIFAR}        & Encoder-I                & 0.5026           &  0.5589          &  0.4975          &  0.5058          & \textbf{0.7267}     \\

                              & Encoder-II               & 0.5068           &  0.6500          &    0.5030        &  0.5147     &  \textbf{0.8200}   \\

                              & Encoder-III              & 0.5041           &  0.7155          &  0.4987          &  0.5103          &  \textbf{0.8868}    \\
\midrule
\multirow{3}{*}{CIFAR-100}    & Encoder-I                & 0.5777           &  0.5549          &  0.5002          &  0.5063          & \textbf{0.6961}     \\
                              & Encoder-II               & 0.4993           &  0.6631          &  0.4973          &  0.5146          & \textbf{0.7922}     \\
                              & Encoder-III              & 0.5049           &  0.6848          &  0.5003          &  0.5106          & \textbf{0.8323}     \\
\midrule
\multirow{3}{*}{TinyImageNet} & Encoder-I                & 0.5051           &  0.5835          &  0.4968          &  0.5019          & \textbf{0.6806}     \\
                              & Encoder-II               & 0.5019           &  0.7467          &  0.5010          &     0.5121       & \textbf{0.7997}    \\
                              & Encoder-III              & 0.5017           &  0.8263          &  0.4974          &  0.5144          & \textbf{0.8343}   \\
\bottomrule
\end{tabular}
}
\label{tab:ASR}
\end{table*}

\mypara{Baseline-C}
In this baseline, the adversary also treats the target encoder directly as a classifier and launches a metric-based membership inference attack~\cite{SM21,YGFJ18}.
Here, we adopt the \textit{entropy} as the metric as it is suitable for the output embedding by the target encoder.
Thus, the adversary can calculate the entropy and search a threshold based on the output embedding of the shadow encoder by feeding the members and non-members of its shadow dataset, and utilize the well-trained binary classifier to infer the members of the target encoder based on the entropy of the output embedding by the target encoder.

\mypara{Baseline-D}
In this baseline, we follow the attack methodology of EncoderMI~\cite{LJQG21}, which aims to attack the pre-trained encoder by contrastive SSL.
EncoderMI considers the input as a member if the model generates similar features for different augmented versions of the input.
In MIM, given an input, we feed it into a pre-trained encoder 10 times. 
Each time we apply randomly scrambled masked patches.
Thus, we obtain 10 embeddings from the encoder calculate the pairwise distances between these embeddings, and get the average distance.
Finally, we consider the input as a member if the average distance is less than a threshold.
The threshold is also derived from the same attack methodology on the shadow dataset and the shadow model.

Here, we emphasize the difference between Baseline-D and our proposed attack from two specific aspects. 
Firstly, we attack against different models: Baseline-D is tailored for contrastive SSL models, whereas we attack masked models. 
Additionally, our insight into membership prediction differs: Baseline-D predicts a member if the encoder produces similar features for the augmented versions of the input. 
In contrast, our attack predicts a member if the reconstructed image is closer to the input.  

\section{Experimental Results}
In this section, we conduct extensive experiments and present the results.

\subsection{Attack Performance}
\mypara{Intuition Verification}
Recall that the key intuition of our attack is that after the masking and reconstruction paradigm of MIM, the reconstructed image should be closer (or less) to the image in the pre-training dataset.
To verify the intuition, we randomly sample 1,000 member images from the pre-training set and 1,000 non-member images from the testing set. 
We then feed these images to the well-pre-trained encoder-decoder to calculate the distance between the reconstructed and original images.
Here we use MSE, a widely used metric, to measure the distance.
As shown in Figure \ref{fig:reconstructed_distance}, we can clearly find that the distance between reconstructed and original images of members is significantly smaller than that of non-members.
These results convincingly verify our key intuition.
The reason behind this is that the encoder-decoder parameters have been optimized for the pre-training set for thousands of iterations, causing the encoder-decoder to over-memorize the pre-training set.

\mypara{Attack ASR Score}
Regarding attack performance, we report \textit{ASR} score in Table \ref{tab:ASR}.
Note that for all baseline attacks, we consider the strongest background knowledge , i.e., the adversary knows the pre-training data distribution, the encoder architecture, and the masking settings.
We can find that only Baseline-B achieves relatively effective attack performance, while Baseline-A, baseline-C, and baseline-D are almost ineffective with ASR scores close to 0.5, a totally random guess.
Encouragingly, in comparison, our attack achieves a much higher \textit{ASR} score than all baseline attacks.
These findings re-confirm our intuition and emphasize the fact that since MIM stems from a new and unique training paradigm i.e., masking and reconstruction, and therefore worth a new attack also based on such paradigms, like ours.

\subsection{Ablation Study}
This section delves into an investigation of how various intrinsic aspects of MIM influence the performance of our attack.
Note that we conduct this study on Encoder-I.

\mypara{The Impact of Encoder's Complexity}
We here investigate the influence of the encoder's complexity.
Recall that Table \ref{tab:model_arches} presents the encoder's complexity, encompassing the number of layers required for its construction and the corresponding parameter sizes.
As evident from the attack's \textit{ASR} score depicted in Table \ref{tab:ASR}, the attack performance increases with the complexity of the encoder.
In other words, the membership leakage of MIM displays a positive correlation with the model's complexity.
This observation can be attributed to the over-parameterization of higher-complexity models, leading to increased memorization of the pre-training set. Consequently, this leads to a widened gap between members and non-members.

\mypara{The Impact of Mask Ratio}
Next, we explore the impact of the mask ratio. 
In particular, we conduct this study by varying the mask ratio from 0.1 to 0.9. 
Figure \ref{vary_mask_ratio} illustrates the \textit{ASR} score regarding the mask ratio.
We can observe that the trend is increasing as the mask ratio rises for both CIFAR-10 and CIFAR-100.
Additionally, the \textit{ASR} score of TinyImageNet remains relatively stable. 
This phenomenon can be attributed to the inherent complexity of TinyImageNet, which exceeds that of CIFAR-10/100. 
Consequently, the model faces greater challenges in generalizing to the testing set (non-members), resulting in a substantial gap between members and non-members.
In general, our findings establish a positive correlation between the mask ratio and the membership leakage of MIM. 
This finding underscores the significance of reconsidering the mask ratio choice, as it challenges the notion that a larger mask ratio is consistently advantageous, as suggested in~\cite{HCXLDG21}. 
Instead, it could render the model more vulnerable to membership leakage. 
Our aim is to stimulate the community to reassess their selection of the mask ratio.

\begin{table}[!t]
\centering
\caption{The impact of mask ratio.}
\label{vary_mask_ratio}
\setlength{\tabcolsep}{5pt}
\scalebox{0.85}{
\begin{tabular}
{c|c|c|c }
\toprule
Mask Ratio &CIFAR-10&CIFAR-100&TinyImageNet\\
\midrule
0.10& 0.5230 & 0.5300 &  0.7432\\
0.25& 0.5290 & 0.5440 &  0.7227\\
0.50& 0.6146 & 0.5940 &  0.7110\\
0.75& 0.7267 & 0.6936 &  0.6905\\
0.90& 0.7637 & 0.7588 &  0.7207\\
\bottomrule
\end{tabular}
}
\end{table}

\mypara{The Impact of Pre-training Epochs}
Lastly, we explore the impact of pre-training epochs. 
We conduct this study by varying the pre-training epochs from 100 to 500. 
Table \ref{table:vary_epochs} show the \textit{ASR} score regarding the pre-training epochs.
We can clearly find the \textit{ASR} score increases in terms of the pre-training epochs, which means the membership leakage of MIM is positively correlated to the pre-training epochs.
The reason behind this observation is that more pre-training epochs make the models over-memorize the pre-training set, resulting in a larger gap between members and non-members.

\begin{table}[!t]
\centering
\caption{The impact of pre-training epochs.}
\label{table:vary_epochs}
\setlength{\tabcolsep}{5pt}
\scalebox{0.85}{
\begin{tabular}
{c|c|c|c }
\toprule
Pre-training Epochs&CIFAR-10&CIFAR-100&TinyImageNet\\
\midrule
100& 0.5205 & 0.5285 &  0.5510\\
200& 0.5620 & 0.5650 &  0.5970\\
300& 0.6311 & 0.6036 &  0.6431\\
400& 0.6756 & 0.6516 &  0.6831\\
500& 0.7272 & 0.6931 &  0.6903\\
\bottomrule
\end{tabular}
}
\end{table}

\subsection{Assumption Relaxation of Adversary Knowledge}\label{relax_assumptions}
Here, we evaluate the performance of our attack by loosening several assumptions made regarding the adversary.

\mypara{Relaxation of Same Decoder Architecture}
As aforementioned, we assume that the shadow decoder has the same architecture as the target decoder.
In this study, we explore the attack performance under the relaxation of this assumption, wherein we vary the architecture of the shadow decoder from 4 to 12 layers, while the target decoder retains 8 layers.
Table \ref{vary_decoder} presents the attack performance across different numbers of shadow decoder layers.
Remarkably, we observe that our attack attains the best performance when both the shadow decoder and target decoder possess an identical architecture, both comprising 8 layers.
Moreover, even when the shadow decoder differs in architecture from the target decoder, our attack retains significant effectiveness, exhibiting only minor degradation from the optimal scenario.
These results indicate that a far more broadly applicable attack scenario is possible. 

\begin{table}[!t]
\centering
\caption{Attack performance under the relaxation of the assumption that the shadow decoder shares the same number of layers as the target decoder.}
\label{vary_decoder}
\setlength{\tabcolsep}{5pt}
\scalebox{0.85}{
\begin{tabular}
{c|c|c|c }
\toprule
Number of Layers&Encoder-I&Encoder-II&Encoder-III\\
\midrule
4& 0.6866 & 0.7822 &  0.8338\\
6& 0.7002 & 0.7887 &  0.8080\\
8& 0.7267 & 0.8200 &  0.8868\\
10& 0.7027 & 0.8003 &  0.7587\\
12& 0.7052 & 0.8293 &  0.7677\\
\bottomrule
\end{tabular}
}
\end{table}

\begin{figure}[t]
    \centering
    \includegraphics[width=0.8\linewidth]{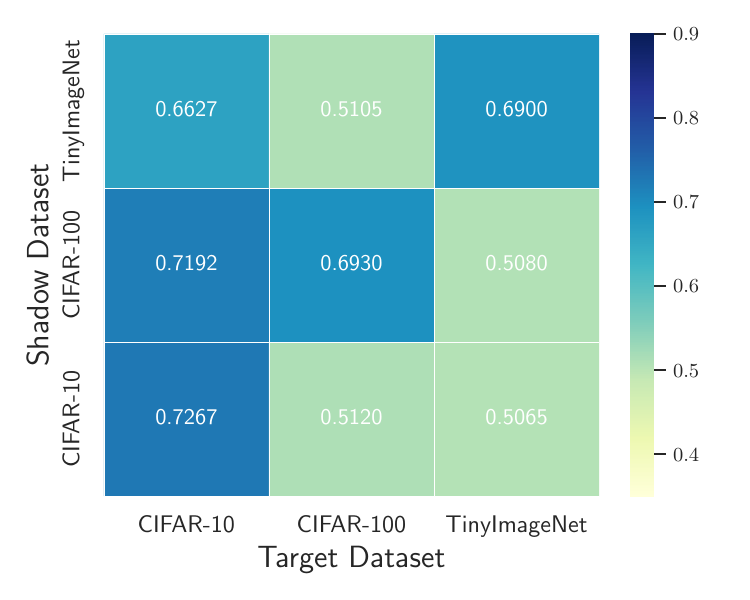}
    \caption{Attack performance under the relaxation of the assumption that the shadow dataset shares the same distribution as the target dataset.}
    \label{relax_dataset}
\end{figure}

\mypara{Relaxation of Dataset Distribution}
Next, we analyze the impact of varying distributions between shadow and target datasets and report the \textit{ASR} score in Figure \ref{relax_dataset}.
Firstly, the highest \textit{ASR} scores consistently emerge along the diagonal, which represents the scenario where the shadow dataset aligns its distribution with the target dataset.
Secondly, it is noteworthy that the adversary still achieves substantial \textit{ASR} scores when the target dataset is CIFAR-10, regardless of the shadow dataset.
More results with the same observation can be found in Appendix Figure \ref{appendix:relax_dataset}.
These findings indicate that even with varying distributions, the adversary can achieve acceptable or remarkable attack performance, particularly if the target dataset holds less complexity or comprehensiveness.

\begin{figure}[t]
    \centering
    \includegraphics[width=0.8\linewidth]{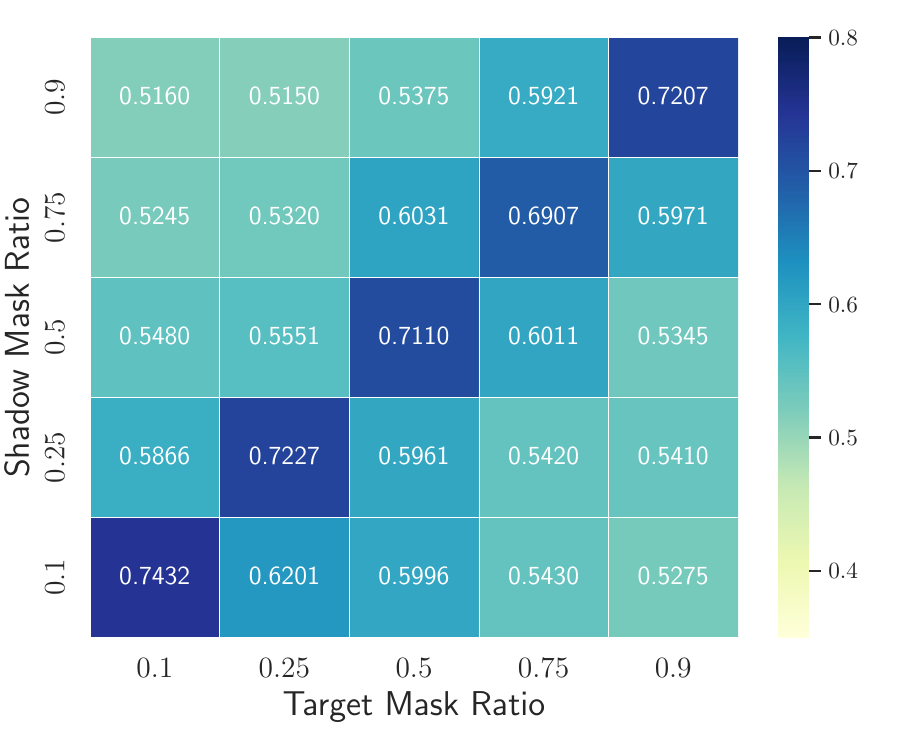}
    \caption{Attack performance under the relaxation of the assumption that the shadow encoder shares the same mask ratio as the target encoder. The pre-training dataset is TinyImageNet.}
    \label{relax_ratio}
\end{figure}

\mypara{Relaxation of Same Mask Ratio}
Here, we relax another assumption that the adversary knows the exact mask ratio adopted by the target encoder.
Figure \ref{relax_ratio} shows the \textit{ASR} score when the mask ratio of the shadow encoder is different from that of the target encoder.
As expected, we can observe that the highest attack \textit{ASR} score is achieved on the diagonal, which is actually the best scenario for the adversary that uses the same mask ratio between the shadow and target encoders.
More results can be found in Appendix Figure \ref{appendix:relax_mask_ratio}.
These results show that the adversary can achieve the best attack performance only if they know the exact mask ratio adopted by the target encoder.
However, we highlight that advanced MIM technologies usually discuss different mask ratio options in their papers~\cite{HCXLDG21,BDW21,DBKWZUDMHGUH21} to get the best results for future tasks. 
This means encoder providers are likely to adopt the best mask ratio. 
Therefore, the adversary could also choose the same one suggested in these advanced techniques.

\begin{table}[]
\centering
\caption{Attack performance under various defenses. The dataset is CIFAR-10.}
\label{Defense}
\scalebox{0.85}{
\begin{tabular}{c|c|c|c }
\toprule
\multirow{2}{*}{Encoder} & \multicolumn{3}{c}{Defense}             \\
                         & None & L2  & Dropout  \\
\midrule
Encoder-I                & 0.7267     &  0.7127  &  0.6432                      \\
Encoder-II               & 0.8200     &  0.8130  &  0.7287                        \\
Encoder-III              & 0.8868     &  0.8820  &  0.7558                \\
\bottomrule
\end{tabular}
}
\end{table}

\section{Defenses}
In this section, we discuss the possible defenses for protecting MIM privacy.

As demonstrated earlier, the pre-trained encoder tends to over-memorize the pre-training dataset, leading to a noticeable difference between members and non-members.
Thus, a straightforward defense is to reduce the extent of over-memorization, such as L2 regularization~\cite{TLGYW18,SZHBFB19}, and Dropout~\cite{SHKSS14}. 
We emphasize that there exist other defenses~\cite{NSH18,JSBZG19}. 
However, they are mainly tailored for classification or generative models.
Thus, we here report the attack performance under these two defenses in Table \ref{Defense}.
We observe that L2 regularization has minimal impact, whereas Dropout demonstrates a comparatively notable enhancement in defensive performance.

Furthermore, it has been observed in previous experiments that the adversary achieves a notably reduced \textit{ASR} score with a different mask ratio. 
This prompts the encoder provider to potentially adopt an unconventional and infrequently used mask ratio during the encoder's training, if they can tolerate a decline in downstream task performance.

Lastly, the encoder provider has the option to obscure details like the pre-training configurations, decoder architectures, and the specifics of the pre-training dataset. Although such opacity would render it more challenging for adversaries to initiate membership inference, it does raise concerns for downstream customers.

\section{Conclusion}
In this work, we pioneer in studying the pre-training data privacy of Masked Image Modeling (MIM) from the perspective of membership inference.
Based on the masking and reconstruction paradigm of MIM, we propose a new attack mechanism unique to MIM, i.e., simulating the masking and reconstruction paradigm and measuring the distance between the reconstructed and original images, which we consider as a member if the distance is smaller than a threshold, and as a non-member otherwise.
We conduct extensive evaluations on various models and benchmark datasets.
Empirical results show that our attack achieves superior performance than existing attack methods.
Further, we conduct an in-depth ablation study on different factors that can guide developers and researchers to be alert to vulnerabilities in the MIM paradigm.
We also investigate which factors of our attack mechanism settings affect the attack performance by relaxing some assumptions.
Lastly, we explore the possible defense methods for MIM against membership inference attacks.

\bibliographystyle{plain}
\bibliography{normal_generated}

\begin{thebibliography}{10}

\bibitem{ABCGL22}
Jianpeng An, Yunhao Bai, Huazhen Chen, Zhongke Gao, and Geert Litjens.
\newblock {Masked Autoencoders Pre-training in Multiple Instance Learning for Whole Slide Image Classification}.
\newblock In {\em {Medical Imaging with Deep Learning (Short Paper) (MIDLS)}}. PMLR, 2022.

\bibitem{BPH22}
Alan Baade, Puyuan Peng, and David Harwath.
\newblock {{MAE-AST:} Masked Autoencoding Audio Spectrogram Transformer}.
\newblock {\em {CoRR abs/2203.16691}}, 2022.

\bibitem{BMAZ22}
Roman Bachmann, David Mizrahi, Andrei Atanov, and Amir Zamir.
\newblock {MultiMAE: Multi-modal Multi-task Masked Autoencoders}.
\newblock {\em {CoRR abs/2204.01678}}, 2022.

\bibitem{BDW21}
Hangbo Bao, Li~Dong, and Furu Wei.
\newblock {BEiT: BERT Pre-Training of Image Transformers}.
\newblock {\em {CoRR abs/2106.08254}}, 2021.

\bibitem{CYZF20}
Dingfan Chen, Ning Yu, Yang Zhang, and Mario Fritz.
\newblock {GAN-Leaks: A Taxonomy of Membership Inference Attacks against Generative Models}.
\newblock In {\em {ACM SIGSAC Conference on Computer and Communications Security (CCS)}}, pages 343--362. ACM, 2020.

\bibitem{CHLZL23}
Joann~Qiongna Chen, Xinlei He, Zheng Li, Yang Zhang, and Zhou Li.
\newblock {A Comprehensive Study of Privacy Risks in Curriculum Learning}.
\newblock {\em {CoRR abs/2310.10124}}, 2023.

\bibitem{CKNH20}
Ting Chen, Simon Kornblith, Mohammad Norouzi, and Geoffrey~E. Hinton.
\newblock {A Simple Framework for Contrastive Learning of Visual Representations}.
\newblock In {\em {International Conference on Machine Learning (ICML)}}, pages 1597--1607. PMLR, 2020.

\bibitem{CDWXMWHLZW22}
Xiaokang Chen, Mingyu Ding, Xiaodi Wang, Ying Xin, Shentong Mo, Yunhao Wang, Shumin Han, Ping Luo, Gang Zeng, and Jingdong Wang.
\newblock {Context Autoencoder for Self-Supervised Representation Learning}.
\newblock {\em {CoRR abs/2202.03026}}, 2022.

\bibitem{CTCP21}
Christopher A.~Choquette Choo, Florian Tram{\`e}r, Nicholas Carlini, and Nicolas Papernot.
\newblock {Label-Only Membership Inference Attacks}.
\newblock In {\em {International Conference on Machine Learning (ICML)}}, pages 1964--1974. PMLR, 2021.

\bibitem{DCAS18}
Luke~Nicholas Darlow, Elliot~J. Crowley, Antreas Antoniou, and Amos~J. Storkey.
\newblock {CINIC-10 is not ImageNet or CIFAR-10}.
\newblock {\em {CoRR abs/1810.03505}}, 2018.

\bibitem{DCLT19}
Jacob Devlin, Ming{-}Wei Chang, Kenton Lee, and Kristina Toutanova.
\newblock {BERT: Pre-training of Deep Bidirectional Transformers for Language Understanding}.
\newblock In {\em {Conference of the North American Chapter of the Association for Computational Linguistics: Human Language Technologies (NAACL-HLT)}}, pages 4171--4186. ACL, 2019.

\bibitem{DBKWZUDMHGUH21}
Alexey Dosovitskiy, Lucas Beyer, Alexander Kolesnikov, Dirk Weissenborn, Xiaohua Zhai, Thomas Unterthiner, Mostafa Dehghani, Matthias Minderer, Georg Heigold, Sylvain Gelly, Jakob Uszkoreit, and Neil Houlsby.
\newblock {An Image is Worth 16x16 Words: Transformers for Image Recognition at Scale}.
\newblock In {\em {International Conference on Learning Representations (ICLR)}}, 2021.

\bibitem{HSLGBZ24}
Ge~Han, Ahmed Salem, Zheng Li, Shanqing Guo, Michael Backes, and Yang Zhang.
\newblock {Detection and Attribution of Models Trained on Generated Data}.
\newblock In {\em {IEEE International Conference on Acoustics, Speech and Signal Processing (ICASSP)}}, pages 4875--4879. IEEE, 2024.

\bibitem{HCXLDG21}
Kaiming He, Xinlei Chen, Saining Xie, Yanghao Li, Piotr Doll{\'{a}}r, and Ross~B. Girshick.
\newblock {Masked Autoencoders Are Scalable Vision Learners}.
\newblock {\em {CoRR abs/2111.06377}}, 2021.

\bibitem{HLXCZ22}
Xinlei He, Zheng Li, Weilin Xu, Cory Cornelius, and Yang Zhang.
\newblock {Membership-Doctor: Comprehensive Assessment of Membership Inference Against Machine Learning Models}.
\newblock {\em {CoRR abs/2208.10445}}, 2022.

\bibitem{HWWBSZ21}
Xinlei He, Rui Wen, Yixin Wu, Michael Backes, Yun Shen, and Yang Zhang.
\newblock {Node-Level Membership Inference Attacks Against Graph Neural Networks}.
\newblock {\em {CoRR abs/2102.05429}}, 2021.

\bibitem{HHB19}
Benjamin Hilprecht, Martin H{\"{a}}rterich, and Daniel Bernau.
\newblock {Monte Carlo and Reconstruction Membership Inference Attacks against Generative Models}.
\newblock {\em {Privacy Enhancing Technologies Symposium}}, 2019.

\bibitem{HHSJCL20}
Lu~Hou, Zhiqi Huang, Lifeng Shang, Xin Jiang, Xiao Chen, and Qun Liu.
\newblock {DynaBERT: Dynamic {BERT} with Adaptive Width and Depth}.
\newblock In {\em {Annual Conference on Neural Information Processing Systems (NeurIPS)}}. NeurIPS, 2020.

\bibitem{HYYBGC21}
Bo~Hui, Yuchen Yang, Haolin Yuan, Philippe Burlina, Neil~Zhenqiang Gong, and Yinzhi Cao.
\newblock {Practical Blind Membership Inference Attack via Differential Comparisons}.
\newblock In {\em {Network and Distributed System Security Symposium (NDSS)}}. Internet Society, 2021.

\bibitem{JLG22}
Jinyuan Jia, Yupei Liu, and Neil~Zhenqiang Gong.
\newblock {BadEncoder: Backdoor Attacks to Pre-trained Encoders in Self-Supervised Learning}.
\newblock In {\em {IEEE Symposium on Security and Privacy (S\&P)}}. IEEE, 2022.

\bibitem{JSBZG19}
Jinyuan Jia, Ahmed Salem, Michael Backes, Yang Zhang, and Neil~Zhenqiang Gong.
\newblock {MemGuard: Defending against Black-Box Membership Inference Attacks via Adversarial Examples}.
\newblock In {\em {ACM SIGSAC Conference on Computer and Communications Security (CCS)}}, pages 259--274. ACM, 2019.

\bibitem{krizhevsky2009learning}
Alex Krizhevsky, Geoffrey Hinton, et~al.
\newblock Learning multiple layers of features from tiny images.
\newblock {\em {}}, 2009.

\bibitem{le2015tiny}
Ya~Le and Xuan Yang.
\newblock Tiny imagenet visual recognition challenge.
\newblock {\em CS 231N}, 7(7):3, 2015.

\bibitem{LLWHYZFZ24}
Hao Li, Zheng Li, Siyuan Wu, Chengrui Hu, Yutong Ye, Min Zhang, Dengguo Feng, and Yang Zhang.
\newblock {SeqMIA: Sequential-Metric Based Membership Inference Attack}.
\newblock {\em {CoRR abs/2407.15098}}, 2024.

\bibitem{LLHYBZ22}
Zheng Li, Yiyong Liu, Xinlei He, Ning Yu, Michael Backes, and Yang Zhang.
\newblock {Auditing Membership Leakages of Multi-Exit Networks}.
\newblock In {\em {ACM SIGSAC Conference on Computer and Communications Security (CCS)}}. ACM, 2022.

\bibitem{LZ21}
Zheng Li and Yang Zhang.
\newblock {Membership Leakage in Label-Only Exposures}.
\newblock In {\em {ACM SIGSAC Conference on Computer and Communications Security (CCS)}}, pages 880--895. ACM, 2021.

\bibitem{LJQG21}
Hongbin Liu, Jinyuan Jia, Wenjie Qu, and Neil~Zhenqiang Gong.
\newblock {EncoderMI: Membership Inference against Pre-trained Encoders in Contrastive Learning}.
\newblock In {\em {ACM SIGSAC Conference on Computer and Communications Security (CCS)}}. ACM, 2021.

\bibitem{NSH18}
Milad Nasr, Reza Shokri, and Amir Houmansadr.
\newblock {Machine Learning with Membership Privacy using Adversarial Regularization}.
\newblock In {\em {ACM SIGSAC Conference on Computer and Communications Security (CCS)}}, pages 634--646. ACM, 2018.

\bibitem{ONK21}
Iyiola~E. Olatunji, Wolfgang Nejdl, and Megha Khosla.
\newblock {Membership Inference Attack on Graph Neural Networks}.
\newblock {\em {CoRR abs/2101.06570}}, 2021.

\bibitem{RKHRGASAMCKS21}
Alec Radford, Jong~Wook Kim, Chris Hallacy, Aditya Ramesh, Gabriel Goh, Sandhini Agarwal, Girish Sastry, Amanda Askell, Pamela Mishkin, Jack Clark, Gretchen Krueger, and Ilya Sutskever.
\newblock {Learning Transferable Visual Models From Natural Language Supervision}.
\newblock In {\em {International Conference on Machine Learning (ICML)}}, pages 8748--8763. PMLR, 2021.

\bibitem{SZHBFB19}
Ahmed Salem, Yang Zhang, Mathias Humbert, Pascal Berrang, Mario Fritz, and Michael Backes.
\newblock {ML-Leaks: Model and Data Independent Membership Inference Attacks and Defenses on Machine Learning Models}.
\newblock In {\em {Network and Distributed System Security Symposium (NDSS)}}. Internet Society, 2019.

\bibitem{SSSS17}
Reza Shokri, Marco Stronati, Congzheng Song, and Vitaly Shmatikov.
\newblock {Membership Inference Attacks Against Machine Learning Models}.
\newblock In {\em {IEEE Symposium on Security and Privacy (S\&P)}}, pages 3--18. IEEE, 2017.

\bibitem{SM21}
Liwei Song and Prateek Mittal.
\newblock {Systematic Evaluation of Privacy Risks of Machine Learning Models}.
\newblock In {\em {USENIX Security Symposium (USENIX Security)}}. USENIX, 2021.

\bibitem{SHKSS14}
Nitish Srivastava, Geoffrey Hinton, Alex Krizhevsky, Ilya Sutskever, and Ruslan Salakhutdinov.
\newblock {Dropout: A Simple Way to Prevent Neural Networks from Overfitting}.
\newblock {\em {Journal of Machine Learning Research}}, 2014.

\bibitem{TLHCCH22}
Qiaoyu Tan, Ninghao Liu, Xiao Huang, Rui Chen, Soo{-}Hyun Choi, and Xia Hu.
\newblock {{MGAE:} Masked Autoencoders for Self-Supervised Learning on Graphs}.
\newblock {\em {CoRR abs/2201.02534}}, 2022.

\bibitem{TSWW22}
Zhan Tong, Yibing Song, Jue Wang, and Limin Wang.
\newblock {VideoMAE: Masked Autoencoders are Data-Efficient Learners for Self-Supervised Video Pre-Training}.
\newblock {\em {CoRR abs/2203.12602}}, 2022.

\bibitem{TLGYW18}
Stacey Truex, Ling Liu, Mehmet~Emre Gursoy, Lei Yu, and Wenqi Wei.
\newblock {Towards Demystifying Membership Inference Attacks}.
\newblock {\em {CoRR abs/1807.09173}}, 2018.

\bibitem{OLV18}
A{\"{a}}ron van~den Oord, Yazhe Li, and Oriol Vinyals.
\newblock {Representation Learning with Contrastive Predictive Coding}.
\newblock {\em {CoRR abs/1807.03748}}, 2018.

\bibitem{WFXWYF21}
Chen Wei, Haoqi Fan, Saining Xie, Chao{-}Yuan Wu, Alan~L. Yuille, and Christoph Feichtenhofer.
\newblock {Masked Feature Prediction for Self-Supervised Visual Pre-Training}.
\newblock {\em {CoRR abs/2112.09133}}, 2021.

\bibitem{WYPY21}
Bang Wu, Xiangwen Yang, Shirui Pan, and Xingliang Yuan.
\newblock {Adapting Membership Inference Attacks to {GNN} for Graph Classification: Approaches and Implications}.
\newblock In {\em {International Conference on Data Mining (ICDM)}}. IEEE, 2021.

\bibitem{WYLBZ22}
Yixin Wu, Ning Yu, Zheng Li, Michael Backes, and Yang Zhang.
\newblock {Membership Inference Attacks Against Text-to-image Generation Models}.
\newblock {\em {CoRR abs/2210.00968}}, 2022.

\bibitem{XZCLBYDH21}
Zhenda Xie, Zheng Zhang, Yue Cao, Yutong Lin, Jianmin Bao, Zhuliang Yao, Qi~Dai, and Han Hu.
\newblock {SimMIM: {A} Simple Framework for Masked Image Modeling}.
\newblock {\em {CoRR abs/2111.09886}}, 2021.

\bibitem{XTLYL20}
Ji~Xin, Raphael Tang, Jaejun Lee, Yaoliang Yu, and Jimmy Lin.
\newblock {DeeBERT: Dynamic Early Exiting for Accelerating {BERT} Inference}.
\newblock In {\em {Annual Meeting of the Association for Computational Linguistics (ACL)}}, pages 2246--2251. ACL, 2020.

\bibitem{XLYCFBZ24}
Yuan Xin, Zheng Li, Ning Yu, Dingfan Chen, Mario Fritz, Michael Backes, and Yang Zhang.
\newblock {Inside the Black Box: Detecting Data Leakage in Pre-trained Language Encoders}.
\newblock In {\em {European Conference on Artificial Intelligence (ECAI)}}, 2024.

\bibitem{YGFJ18}
Samuel Yeom, Irene Giacomelli, Matt Fredrikson, and Somesh Jha.
\newblock {Privacy Risk in Machine Learning: Analyzing the Connection to Overfitting}.
\newblock In {\em {IEEE Computer Security Foundations Symposium (CSF)}}, pages 268--282. IEEE, 2018.

\bibitem{ZZSYZK22}
Chaoning Zhang, Chenshuang Zhang, Junha Song, John Seon~Keun Yi, Kang Zhang, and In~So Kweon.
\newblock {A Survey on Masked Autoencoder for Self-supervised Learning in Vision and Beyond}.
\newblock {\em {CoRR abs/2208.00173}}, 2022.

\bibitem{ZXGMXW20}
Wangchunshu Zhou, Canwen Xu, Tao Ge, Julian~J. McAuley, Ke~Xu, and Furu Wei.
\newblock {{BERT} Loses Patience: Fast and Robust Inference with Early Exit}.
\newblock In {\em {Annual Conference on Neural Information Processing Systems (NeurIPS)}}. NeurIPS, 2020.

\end{thebibliography}

\clearpage
\appendix

\section{Appendix}
\subsection{Datasets}
Here's a brief introduction to each dataset:

\mypara{MNIST} MNIST is a gray image dataset consisting of handwritten digits. It contains 60,000 training images and 10,000 test images, with each image representing a single digit from 0 to 9.

\mypara{CIFAR-10} CIFAR-10 is a dataset composed of 60,000 color images divided into 10 classes. Each image in CIFAR-10 belongs to one of the following categories: airplane, automobile, bird, cat, deer, dog, frog, horse, ship, or truck.

\mypara{TinyImageNet-100}: TinyImageNet-100 is a subset of the larger ImageNet dataset, containing 100 object classes with 500 training images and 50 validation images per class.

\subsection{Threshold Setting}
As aforementioned, one key stage of our proposed attack is to search for an optimal threshold based on the shadow encoder and encoder.
This searched threshold can be used to differentiate members from non-members of the target model.
Table \ref{table:threshold} shows the threshold we searched for in our evaluation.
\begin{table}[!h]
\centering
\caption{The distance score threshold.}
\label{table:threshold}
\setlength{\tabcolsep}{5pt}
\scalebox{0.8}{
\begin{tabular}
{c|c|c|c }
\toprule
Threshold  &Encoder-I&Encoder-II&Encoder-III\\
\midrule
CIFAR-10& 0.0022 & 0.0017 &  0.0009\\
CIFAR-100& 0.0028 & 0.0019 &  0.0009\\
TinyImageNet& 0.0088 & 0.0068 &  0.0042\\
\bottomrule
\end{tabular}
}
\end{table}

\begin{figure*}[!t]
\centering
\begin{subfigure}{0.65\columnwidth}
\includegraphics[width=\columnwidth]{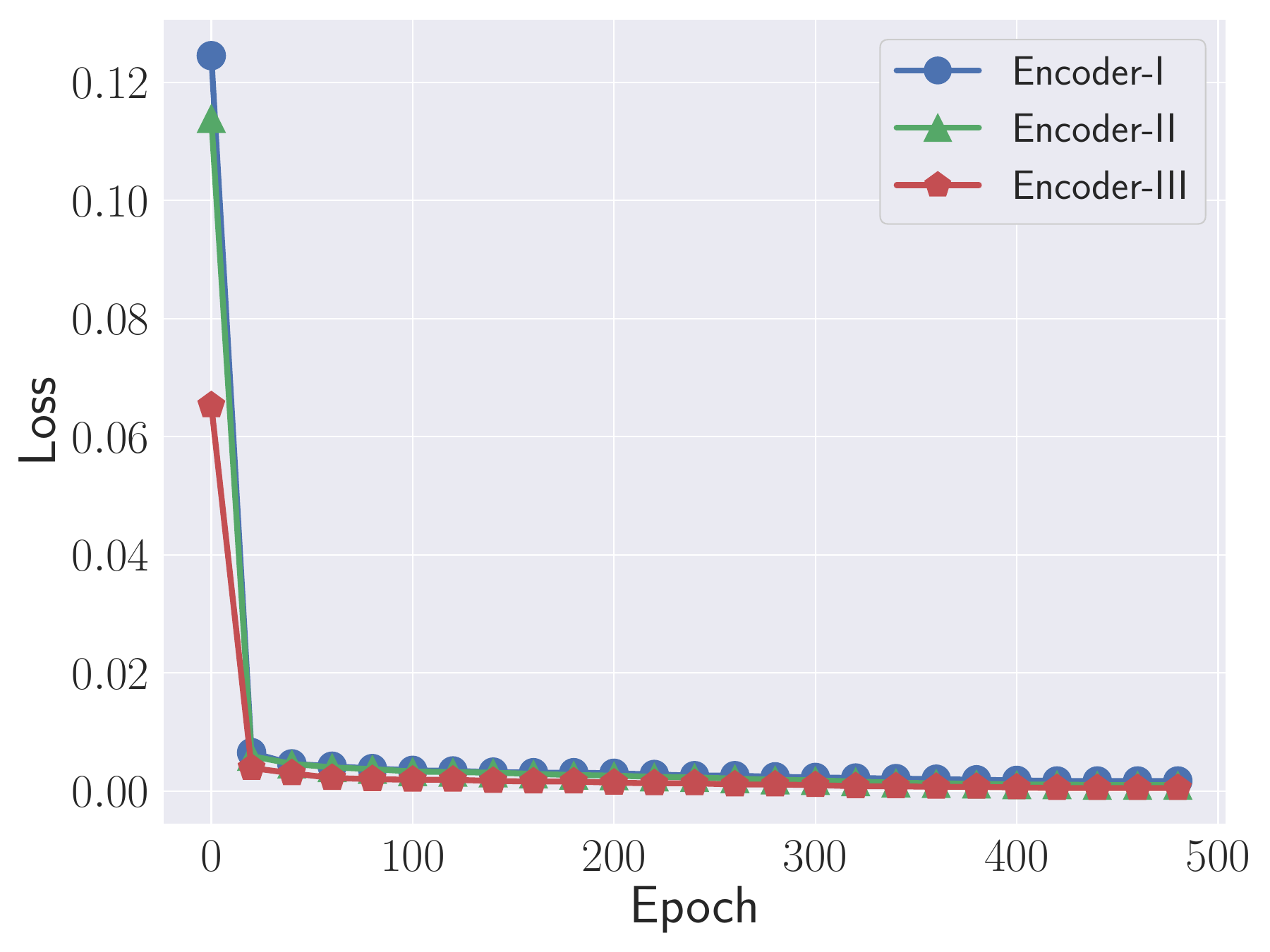}
\caption{CIFAR-10}
\end{subfigure}
\begin{subfigure}{0.65\columnwidth}
\includegraphics[width=\columnwidth]{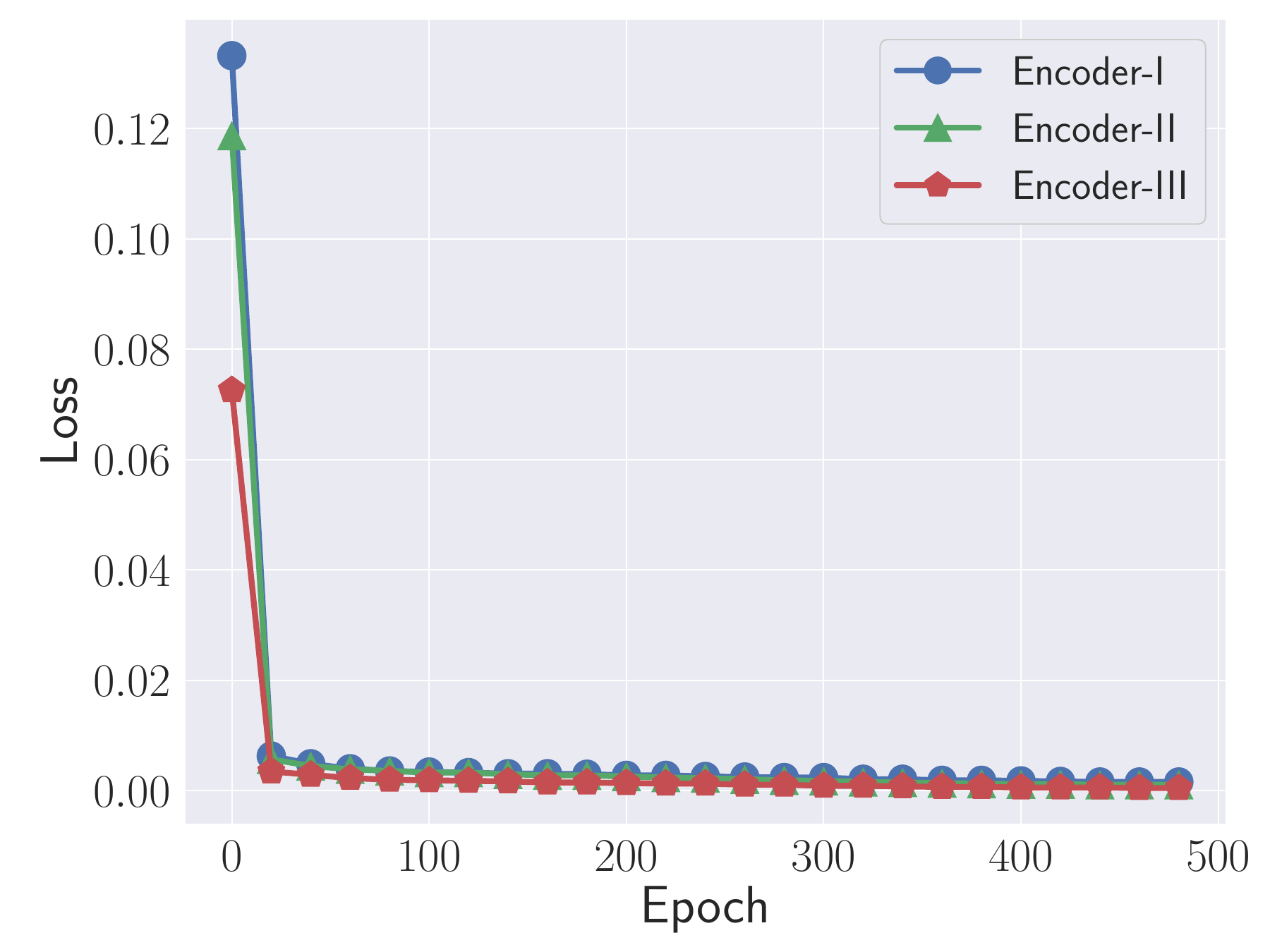}
\caption{CIFAR-100}
\end{subfigure}
\begin{subfigure}{0.65\columnwidth}
\includegraphics[width=\columnwidth]{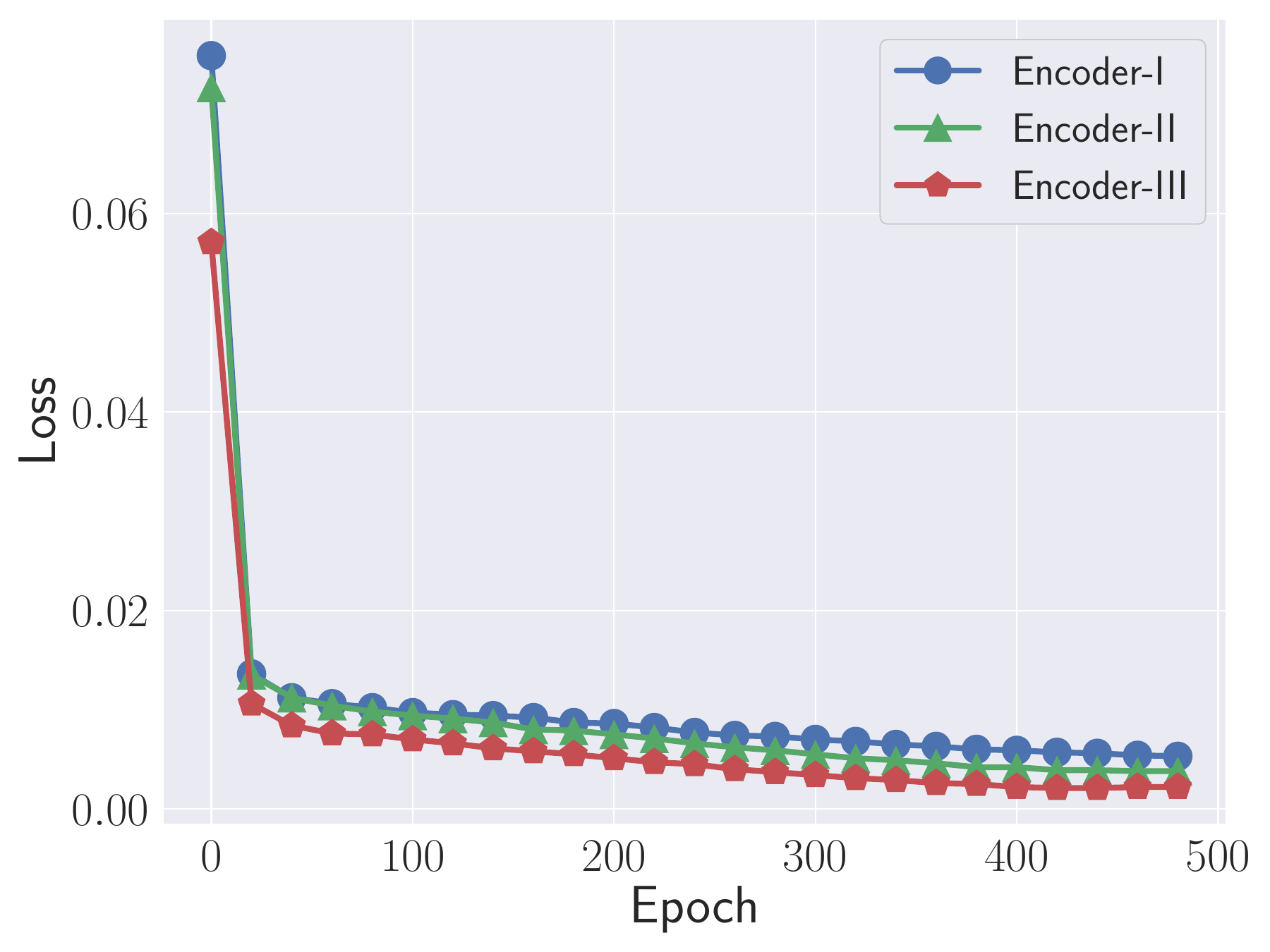}
\caption{TinyImageNet}
\end{subfigure}
\caption{The trend of loss in terms of the  pre-training epochs.}
\label{fig:training_loss}
\end{figure*}

\begin{figure*}[!t]
\centering
\begin{subfigure}{0.9\columnwidth}
\includegraphics[width=\columnwidth]{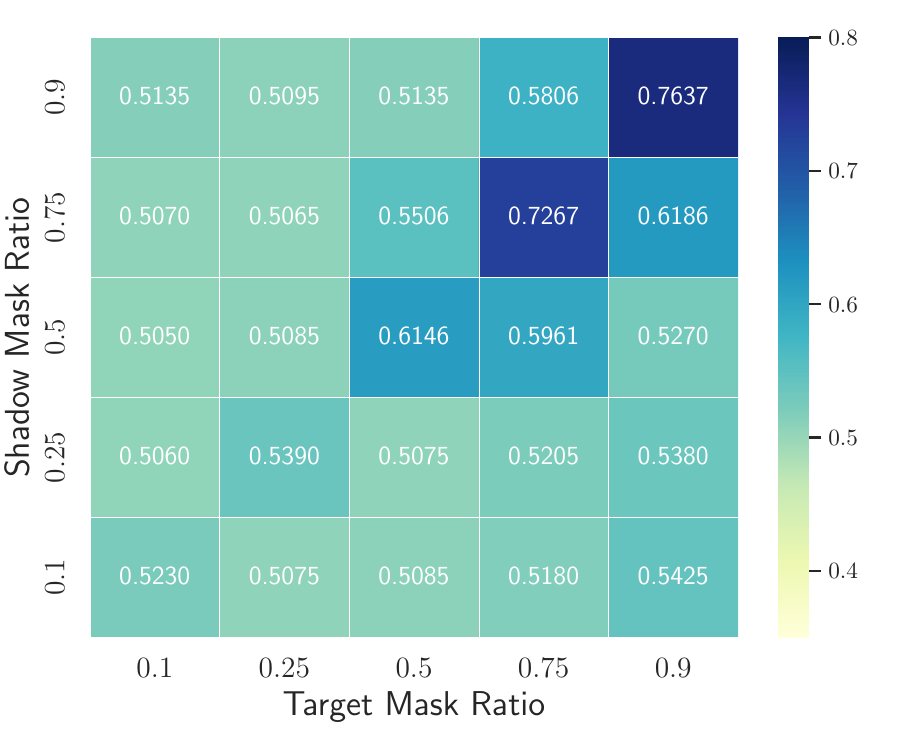}
\caption{CIFAR-10}
\end{subfigure}
\begin{subfigure}{0.9\columnwidth}
\includegraphics[width=\columnwidth]{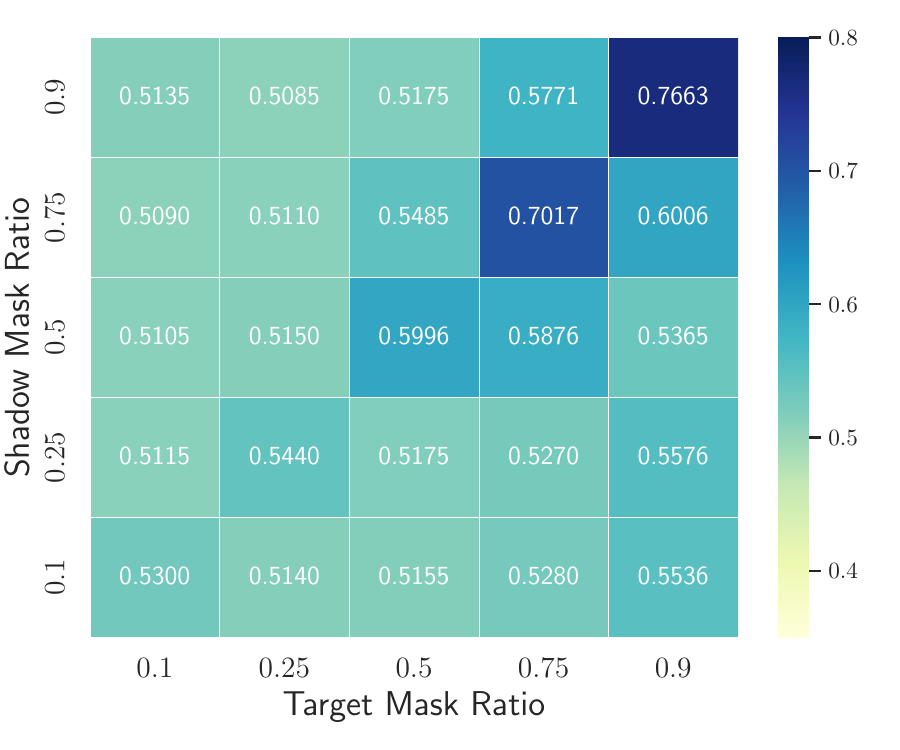}
\caption{CIFAR-100}
\end{subfigure}
\caption{Attack performance under the relaxation of the assumption that the shadow encoder shares the same mask ratio as the target encoder.}
\label{appendix:relax_mask_ratio}
\end{figure*}

\begin{figure*}[!t]
\centering
\begin{subfigure}{0.8\columnwidth}
\includegraphics[width=\columnwidth]{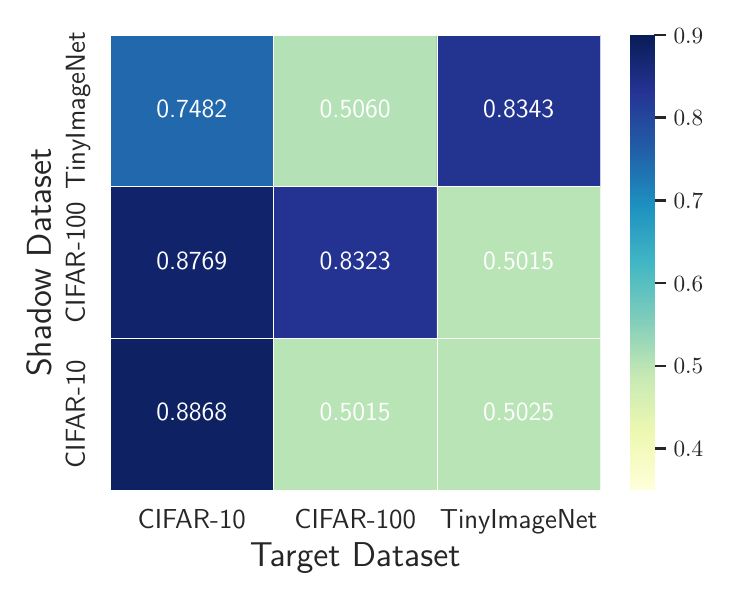}
\caption{Encoder-II}
\end{subfigure}
\begin{subfigure}{0.8\columnwidth}
\includegraphics[width=\columnwidth]{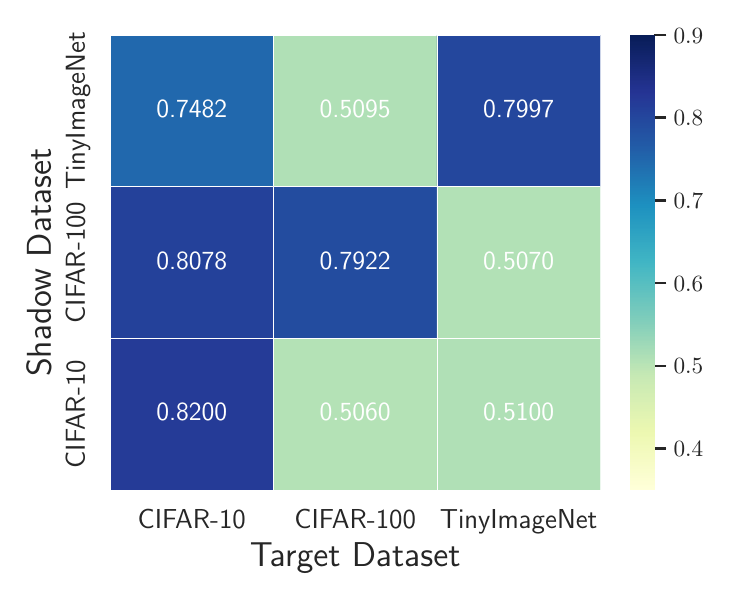}
\caption{Encoder-III}
\end{subfigure}
\caption{Attack performance under the relaxation of the assumption that the shadow dataset shares the same distribution as the target dataset.}
\label{appendix:relax_dataset}
\end{figure*}

\end{document}